\documentclass[Crown]{sagej}
\usepackage{natbib}
\usepackage{float}
\usepackage{amsmath,amsthm,amssymb}

\newcommand{\vect}[1]{\mathbf{#1}}

\newtheorem{remark}{Remark}

\begin{document}
\runninghead{B\"{a}umer et al.}
\title{A general framework for the asymptotic analysis of moist atmospheric flows}
\author{Daniel B\"{a}umer\affilnum{1}, Rupert Klein\affilnum{2}, Norbert J. Mauser\affilnum{1}}
\affiliation{\affilnum{1}Research platform MMM, c/o Fakult\"{a}t f\"{u}r Mathematik, Universit\"{a}t Wien, Vienna, Austria \\
\affilnum{2}FB f\"{u}r Mathematik und Informatik, Freie Universit\"{a}t Berlin, Berlin, Germany}
\corrauth{Daniel B\"{a}umer (Research platform MMM, c/o Fakult\"{a}t f\"{u}r Mathematik, Universit\"{a}t Wien. Oskar-Morgenstern-Platz 1-3, A-1090 Wien)}
\email{daniel.baeumer@univie.ac.at}
\begin{abstract}
We deal with asymptotic analysis for the derivation of partial differential equation models for geophysical flows in the earth's atmosphere with moist process closures, and we study their mathematical properties. Starting with the Navier-Stokes equations for dry air, we put the seminal papers of Klein, Majda et al.~in a unified context and then discuss the  appropriate extension to moist air. In particular, we deal with the scale-independent distinguished limit for the universal parameters of atmospheric motion for moist air, with the Clausius-Clapeyron relation that links saturation vapor pressure and air temperature, and with the mathematical formulation of phase changes associated with cloud formation and rain production. We conclude with a discussion of the \emph{precipitating quasi-geostrophic} (PQG) models introduced by Smith \& Stechmann. Our intent is, on the one hand, to convey the problems arising at the modeling stage to mathematicians; on the other hand, we want to present the relevant mathematical methods and results to meteorologists.
\end{abstract}
\maketitle

\section{Introduction}
In this paper, we discuss recent developments concerning challenges associated with the derivation and mathematical analysis of reduced models for atmospheric motion with moist process closures. Specifically, we focus on the following topics:
\begin{enumerate}
    \item The choice of a distinguished limit: as laid out in \citep{klein2010}, the rich palette of reduced models used in meteorology can be viewed as an ensemble of scale-specific models that are embedded in \emph{one} scale-independent distinguished limit for the universal parameters of atmospheric motion on our planet. Said distinguished limit, however, does not incorporate the thermodynamic parameters of moist air. This raises the question, crucially important for realistic models, how its systematic extension to that end may be achieved. Here, we base our presentation on the discussion in \citep{hittmeir2018}, which shows that it is a difficult task even just to reconcile formal consistency and physical realism in this endeavor.
    \item The Clausius-Clapeyron relation: the consistency issues just mentioned are mainly due to the relationship between saturation vapor pressure and air temperature encoded in the famous Clausius-Clapeyron equation \citep{bohren1998}, which was systematically studied for the first time in the context of formal asymptotic analysis by \citet{klein2006}. We compare and contrast several approaches to resolving this issue. Furthermore, we highlight the significant ways in which the mathematical structure of the reduced moisture equations changes, depending on the chosen approach.
    \item The impact of phase changes: one of the most important features of bulk models for cloudy, precipitating air is the presence of closures for the phase transitions associated with cloud formation and rain production. Viewed through a mathematical lens, the first-order discontinuities that occur in these closures provide an interesting challenge already when investigating the full, unapproximated equations of geophysical fluid dynamics \citep{hittmeir2017,hittmeir2020,hittmeir2023,doppler2024}. In reduced models, the effect of such discontinuities can lead to even more drastic changes in the analytical structure of moist extensions of preexisting models. Here, the precipitating quasi-geostrophic (PQG) model family originally due to \citet{smith2017} serve as a striking example: the elliptic inversion equation for the pressure perturbation in the potential vorticity formulation of those model equations becomes nonlinear by the inclusion of moisture species, indicating a significant departure from the well-known dry QG dynamics. Once more, we present two contrasting approaches to the moisture parametrization problem, one of which was recently proposed by two of the authors of this article in \citep{baeumer2023}.
\end{enumerate}
We conclude with an outlook on open questions and areas for future research both on the modeling and the rigorous analysis front.
\section{Model hierarchies in atmospheric fluid dynamics: a general framework}
\subsection{Preliminaries}
As it is our primary goal to delineate the physical and mathematical conditions that must be met in order to successfully incorporate moisture into asymptotic models for atmospheric dynamics, we first need to identify the independent physical dimensions of the system's unknowns. In any non-isothermal, compressible fluid, these comprise a velocity field, pressure, density and temperature at a minimum, which means that at least the four independent dimensions length, time, mass and temperature are involved. For the description of geophysical fluids, these turn out to be sufficient. When investigating a specific geophysical flow phenomenon, one can then conduct a preliminary formal analysis in the following familiar fashion: having identified all of its (dimensional) parameters, derive a representative set of independent dimensionless parameters, translate heuristic notions of ``large'' and ``small'' to limit processes and study how the governing equations change in the formal limit(s).

In this work, before discussing scale-specific models, we want to make clear how such a dimensional analysis can be applied in a far more general context: the approach pioneered by and outlined in \citep{klein2010} is to first single out a set of parameters, called \emph{universal characteristics,} that do not depend on the choice of length and timescales appropriate for any particular atmospheric flow. The independent nondimensional parameters that can be derived from said universal characteristics can then be assigned asymptotic scalings to obtain a \emph{scale-independent distinguished limit.} This limit, when chosen properly and supplemented by suitable length - and timescales for some concrete meteorological event, can be used to retrieve already known reduced models, but, more importantly, it clarifies their respective standing within the atmospheric model hierarchy. Establishing a scale-independent distinguished limit thus greatly facilitates
\begin{enumerate}
    \item[(i)] the comparison of different reduced models and
    \item[(ii)] the study of interactions between such models across multiple scales.
\end{enumerate}
\paragraph{The necessity of distinguished limits:} It is well known that asymptotic expansions with respect to multiple small parameters in general yield nonunique limits. A particularly simple, yet illuminating example is provided by the linear oscillator with small mass and damping \citep{klein2010}. In technical terms, this is so because the system's differential operator is not Fr\'{e}chet-differentiable at the origin of the parameter space, rendering the limit dependent on the path toward the origin taken in said space. For this reason, we need to restrict ourselves to \emph{coupled} limit processes dependent on only one generic small parameter $\epsilon$, in other words, a distinguished limit.
\subsection{Universal characteristics of atmospheric motion}
When it comes to the mathematical description of \emph{dry} air flows in the earth's atmosphere, there is general agreement on the equations to be used as a point of departure. These are the compressible Navier-Stokes equations with gravity for a heat-conducting ideal gas in a rotating frame, considered in a thin shell around an (approximately) spherical planet \citep{pedlosky1987,vallis2017}. In coordinate-free form, they read
\begin{subequations} \label{GovEqs_dry_unapprox}
    \begin{align}
        \partial_t (\rho \vect{v}) + \nabla \cdot (\rho \vect{v} \otimes \vect{v}) + 2 (\vect{\Omega} \times \rho \vect{v}) + \nabla p &= \nabla \cdot \sigma + \rho \nabla \Phi \\
        \partial_t \rho + \nabla \cdot (\rho \vect{v}) &= 0 \\
        \partial_t (\rho e) + \nabla \cdot (\rho e \vect{v} + p \vect{v}) &= \nabla \cdot (\sigma \vect{v}) - \nabla \cdot \vect{j} + \rho \vect{v} \cdot \nabla \Phi + \rho Q.
    \end{align}
\end{subequations}
Here, $\vect{v}$ denotes the mass-averaged fluid velocity, while $p$ and $\rho$ stand for pressure and density, respectively; $\vect{\Omega}$ is the earth rotation vector, $\Phi$ the geopotential, incorporating gravitational and centrifugal potential energy, $Q$ is a stand-in for any and all external heat sources, and the following constitutive relations hold:
\begin{itemize}
    \item The viscous stress tensor of a Newtonian fluid, denoted by $\sigma$, reads
    \begin{equation} \label{viscous_stress}
        \sigma = \mu \left(\nabla \vect{v} + \nabla \vect{v}^{\mathrm{T}}\right) + (\zeta - \frac{2}{3}\mu) (\nabla \cdot \vect{v}) I.
    \end{equation}
    Here, $I$ is the identity matrix, and $\mu$ and $\zeta$ denote the viscosity coefficients of the fluid.
    \item The ideal gas law
    \begin{equation}
        p = R_d \rho T,
    \end{equation}
    where $R_d$ denotes the gas constant of dry air and $T$ the temperature of the fluid.
    \item The sum of kinetic and inner energy per unit mass $e$ is given by
    \begin{equation}
        e = \frac{\vect{v}^2}{2} + c_{\text{vd}} T,
    \end{equation}
    with the heat capacity of dry air at constant volume $c_{\text{vd}}$.
    \item The heat flux density $\vect{j}$ obeys Fourier's law and thus reads
    \begin{equation} \label{heat_flux}
        \vect{j} = -\lambda \nabla T,
    \end{equation}
    with the thermal conductivity $\lambda$.
\end{itemize}
In meteorological applications, a number of simplifications typically is assumed from the outset: first of all, the impact of molecular viscosity on meteorological timescales is generally viewed as negligible. Therefore, the terms involving the viscous stress tensor \eqref{viscous_stress} in the momentum and energy equations are typically dropped and replaced by a generic closure term that primarily represents boundary layer friction or turbulence. By the same token, heat conduction proceeds far too slowly to affect the local heat budget in atmospheric motions, and consequently, \eqref{heat_flux} is either ignored altogether or replaced by a closure for turbulent mixing. It remains to discuss the geopotential $\Phi$ and the geometry of planet earth, where quite subtle issues arise: the centrifugal force, while small compared to gravity, \emph{does} make a non-negligible contribution in the plane perpendicular to the earth's gravity vector. As explained, e.g., in chapter 2 of \citet{vallis2017}, naively choosing the latter as the ``vertical'' direction would then have the undesirable consequence of introducing the centrifugal force as a dominant term in the ``horizontal'' momentum equations, thus obscuring the balance between horizontal pressure gradient and Coriolis force that is characteristic of large-scale weather systems in the middle latitudes. The customary way out of this dilemma is the following: one introduces an \emph{effective gravity} as the sum of gravitational and centrifugal forces and defines the ``vertical'' direction accordingly. This yields horizontal surfaces as those of constant geopotential. To be sure, these are not perfect spheres, but due to the fact that the earth is an \emph{oblate spheroid}, i.e., it is flattened at the poles and bulging at the equator, they are very nearly parallel to the actual surface of our planet. Neglecting the (small) deviation from sphericity from that point on is traditional - see \citet{constantin2021} for a detailed investigation that incorporates said deviation in an asymptotic ansatz.
\paragraph{Metric terms in common approximations:} When studying models for atmospheric dynamics on length scales significantly smaller than the earth's diameter, one can simplify the equations of motion further by approximating the spherical metric by a \emph{flat} one (at leading order). This corresponds to a \emph{tangent plane approximation}, which could in principle be justified within the asymptotic scheme. For simplicity of exposition, we will assume it here from the outset, since it is known to be consistent with all models investigated in this article, and the general scaling considerations to follow do not depend on the components of the metric tensor. Bearing all of the above in mind, one can now rewrite the simplified Eqs.~\eqref{GovEqs_dry_unapprox} in pseudo-cartesian coordinates, in the form
\begin{subequations} \label{GovEqs_dry}
\begin{align}
D_t \vect{u}+2(\vect{\Omega}\times\vect{v})_\parallel+\frac{1}{\rho}\nabla_\parallel p &= \mathcal{D}_{\vect{u}} \\
D_t w +2(\vect{\Omega}\times\vect{v})_\perp + \frac{1}{\rho} \partial_z p &= -g + \mathcal{D}_w\\
\partial_t \rho + \nabla\cdot(\rho \vect{v}) &= 0 \\
c_{\text{pd}} D_t \ln\left(\frac{\theta}{\theta_{\text{ref}}}\right) &= \mathcal{D}_\theta + Q. \label{GovEqs_dry_temp}
\end{align}
\end{subequations}
Here, $\vect{v} = (\vect{u},w)^\mathrm{T}$, where $\vect{u}$ denotes the horizontal velocity field in the tangent plane approximation; we introduced the material derivative $D_t$, defined as
\begin{equation}
    D_t := (\partial_t + \nabla \cdot \vect{v})
\end{equation}
and rewrote the energy equation in terms of the \emph{potential temperature}
\begin{equation}
    \theta = T \left(\frac{p}{p_{\text{ref}}}\right)^{\frac{\gamma - 1}{\gamma}}
\end{equation}
($\gamma$ is the isentropic exponent of dry air). This quantity is of central importance, because its logarithm, multiplied by the heat capacity of dry air at constant pressure $c_{\text{pd}}$, yields the \emph{specific entropy} of a dry air parcel \citep{bohren1998}. Therefore, \eqref{GovEqs_dry_temp} is a transport equation for entropy, and the term $Q$, accordingly, now represents sources of entropy (we reused the same letter for simplicity, since we will not refer to \eqref{GovEqs_dry_unapprox} from here on out). The terms $\mathcal{D}_\star$ on the respective right-hand sides represent closures for turbulent friction, diffusion and mixing, while $g$ represents the acceleration of gravity (which, strictly speaking, incorporates the centrifugal force - see the discussion above).

Atmospheric flows occur on a wide variety of spatial and temporal scales, and the derivation of models of reduced complexity to reveal the characteristics of scale-specific phenomena has long been a cornerstone of theoretical meteorology. Traditionally, such models are derived from the governing equations by a set of assumptions tailored to the problem at hand, followed by careful scale analysis and the identification of dominant terms in the governing equations. This is a standard modeling strategy that has led to many successes, but it has one major drawback: problem-by-problem modeling does not enable the study of \emph{interactions across multiple scales.} Such studies constitute the natural next step when one tries to go beyond single-scale models, and it would therefore be of great interest to establish a framework that allows for the systematic derivation of reduced models by formal asymptotics on \emph{all} viable length and timescales, including, but not limited to, those already recognized by the meteorological community. As already mentioned, this was achieved in the context of dry air flows by \citet{klein2010}, whose methodology we explain more thoroughly in the following:

Notwithstanding the complexity and unpredictability of the dynamics of our planet's atmosphere, one can identify a number of relevant characteristic quantities that are ``universal'' in the sense that they set conditions that remain essentially unchanged for \emph{all} atmospheric flows, listed in Table \ref{tab:univ_char}.
\begin{table}[H]
\centering
\begin{tabular}{l|r}
   Earth's radius   & $a \sim 6 \times 10^6\;\text{m}$ \\
   Earth's rotation rate  & $\Omega \sim 10^{-4}\;\text{s}^{-1}$ \\
   Acceleration of gravity & $g \sim 9.81\;\text{m s}^{-2}$ \\
   Sea-level pressure & $p_{\text{ref}} \sim 10^5\;\text{kg m}^{-1}\;\text{s}^{-2}$ \\
   Water-freezing temperature & $T_{\text{ref}} \sim 273$ K \\
   Tropospheric vertical potential temperature difference & $\Delta \Theta \sim 40$ K \\
   Dry air gas constant & $R_d = 287\;\text{m}^{2}\;\text{s}^{-2}\;\text{K}^{-1}$ \\
   Isentropic exponent of dry air & $\gamma = 1.4$ \\
\end{tabular}
\caption{Universal characteristics of atmospheric motion}
\label{tab:univ_char}
\end{table}
Among these eight parameters, six where already introduced through the governing equations and the constitutive relations. The earth's radius $a$ is necessary to define the spatial domain and consequently also part of the mathematical formulation. Thus, only $\Delta \Theta$ remains as an empirically given quantity - however, if we imagine a setting in which an upper boundary is assumed at the tropopause, this temperature difference could also be derived from the associated boundary condition. We further mention that the equator-to-pole potential temperature difference at fixed altitude is of the same order and could therefore as well be used to define $\Delta \Theta$. Looking ahead to the incorporation of moist quantities, we should also note that the temperature increase that would be incurred if all water vapor in, say, a tropical region condensed out also is of the same order of magnitude. Thus, $\Delta \Theta$ can straightforwardly be derived from the system's parameters if moist processes are included. As far as scaling considerations go, the isentropic exponent of dry air $\gamma$ is already nondimensional. Since the remaining seven involve the four physical dimensions of length, time, mass and temperature, systematic nondimensionalization yields three independent dimensionless parameters, the choice of which should be guided by meteorological considerations. Here, it is convenient to introduce a number of important derived quantities that are summarized in Table \ref{tab:derived_quantities}.
\begin{table}[H]
    \centering
    \begin{tabular}{l|r}
    Sea-level air density     & $\rho_{\text{ref}} = p_{\text{ref}}/(R_d T_{\text{ref}}) \sim 1.25\;\text{kg m}^{-3}$ \\
    Density scale height     & $h_{\text{sc}} = \gamma p_{\text{ref}}/(g \rho_{\text{ref}}) \sim 11$ km \\
    Speed of sound & $c_{\text{ref}} = \sqrt{\gamma p_{\text{ref}}/\rho_{\text{ref}}} \sim 330\;\text{m s}^{-1}$ \\
    Internal wave speed & $c_{\text{int}} = \sqrt{g h_{\text{sc}} \frac{\Delta \Theta}{T_{\text{ref}}}} \sim 110\;\text{m s}^{-1}$ \\
    Thermal wind velocity & $u_{\text{ref}} = \frac{2}{\pi} \frac{g h_{\text{sc}}}{\Omega a} \frac{\Delta \Theta}{T_{\text{ref}}} \sim 12\;\text{m s}^{-1}$
    \end{tabular}
    \caption{Quantities derived from universal characteristics}
    \label{tab:derived_quantities}
\end{table}
With these in hand, we can write the independent dimensionless parameters chosen by Klein [2010] in the form
\begin{align}
    \Pi_1 &= \frac{h_{\text{sc}}}{a} \sim 1.6 \times 10^{-3}, \nonumber \\
    \Pi_2 &= \frac{\Delta \Theta}{T_{\text{ref}}} \sim 1.5 \times 10^{-1}, \\
    \Pi_3 &= \frac{c_{\text{ref}}}{\Omega a} \sim 4.7 \times 10^{-1}. \nonumber
\end{align}
The parameter $\Pi_1$ relates the vertical extent of the troposphere (which is almost exclusively responsible for weather phenomena) to the earth's radius; $\Pi_2$ gives an estimate for the relative strength of atmospheric temperature variations, while $\Pi_3$ compares a characteristic signal speed to the rotation speed of our planet. Proceeding further, the asymptotic rescaling of these parameters is supposed to serve as the common thread uniting the multitude of reduced models that can be derived from the governing equations. Due to the nonuniqueness of multi-parameter expansions, we cannot treat $\Pi_1$, $\Pi_2$ and $\Pi_3$ as independent parameters when passing to an abstract asymptotic setting, and we therefore need to find a suitable distinguished limit. As for the generic small parameter $\epsilon$, it is traditionally assumed in meteorological studies that it corresponds to a numerical value of $\sim 1/10$, a natural choice for a ``scale separation factor'' (see \citet{klein2010} for more background on scale separation). As shown in the cited article, the choice
\begin{equation} \label{dist_lim_dry}
    \begin{array}{ccc}
    \Pi_1 = c_1 \epsilon^3,     & \Pi_2 = c_2 \epsilon, & \Pi_3 = c_3 \sqrt{\epsilon},
    \end{array}
\end{equation}
where the prefactors of the form $c_i$ are to be understood as bounded quantities in the limit $\epsilon \rightarrow 0$, enables the straightforward derivation of almost all of the classical reduced models utilized by meteorologists, and it thus provides a strong and flexible foundation for future investigations, in particular of the aforementioned interactions across multiple scales.
\subsection{Distinguished limits for moist air and the Clausius-Clapeyron relation}
Now, we want to explore possibilities to extend the distinguished limit described above to include the physical parameters that can be regarded as universal when investigating a mixture of dry air, water vapor and liquid water. In doing so, we do not immediately discuss the difficult issue of the incorporation of cloud microphysics, first highlighting the role of the fundamental \emph{Clausius-Clapeyron relation} \citep{bohren1998} in the study of phase changes; the next section will be devoted to viable parameterizations of processes such as condensation and evaporation in the context of formal asymptotics for the dynamics of a moist atmosphere.

First, we of course need to determine a set of governing equations that is suited to serve as a common point of departure for the derivation of a wide variety of moist flow phenomena. Here, we opt for the state-of-the-art system used in \citep{hittmeir2018,baeumer2023,baeumer2025}, which incorporates fairly detailed moist thermodynamics and bulk microphysics closures for transitions between the various moisture species in the spirit of \citet{kessler1995}:
\begin{subequations}\label{GovEqs}
\begin{align}
\label{GovEqs_momentum_hor}
D_t\vect{u}+2(\vect{\Omega}\times\vect{v})_\parallel+\frac{1}{\rho}\nabla_\parallel p
  & = \frac{\rho_d}{\rho}q_r V_r\partial_z\vect{u}+\mathcal{D}_{\vect{u}}, 
    \\
\label{GovEqs_momentum_vert}
D_tw+2(\vect{\Omega}\times\vect{v})_\perp+\frac{1}{\rho}\partial_z p
  & = -g+\frac{\rho_d}{\rho}q_r V_r\partial_z w+\mathcal{D}_w, 
    \\
\label{GovEqs_mass}
D_t\rho_d+\rho_d(\nabla\cdot{\vect{v}})
  & = 0, 
    \\
\label{GovEqs_temp}
C D_t\ln\left(\frac{\theta}{\theta_{\text{ref}}}\right)+\Sigma D_t\ln\left(\frac{p}{p_{\text{ref}}}\right) \nonumber \\
+\frac{L(T)}{T}D_tq_v
  & = c_lV_rq_r\left(\partial_z\ln\left(\frac{\theta}{\theta_{\text{ref}}}\right)+\frac{R_d}{c_{\text{pd}}}\partial_z\ln\left(\frac{p}{p_{\text{ref}}}\right)\right) \nonumber \\
  & +Q+\mathcal{D}_\theta, 
    \\
\label{GovEqs_qv}
D_tq_v
  & = S_{\text{ev}}-S_{\text{cd}}+\mathcal{D}_v, 
    \\
\label{GovEqs_qc}
D_tq_c
  & = S_{\text{cd}}-S_{\text{cr}}-S_{\text{ac}}+\mathcal{D}_c, 
    \\
\label{GovEqs_qr}
D_tq_r-\frac{1}{\rho_d}\partial_z(\rho_dV_rq_r)
  & = S_{\text{cr}}+S_{\text{ac}}-S_{\text{ev}}+\mathcal{D}_r,
\end{align}
\end{subequations}
with the supplementary relations
\begin{subequations}\label{GovEqs_Supplement}
\begin{align}
\label{GovEqs_Supplement-a}
\rho
  & =\rho_d(1+q_v+q_c+q_r), 
    \\
\label{GovEqs_Supplement-b}
p
  & = R_d\rho_dT(1+\frac{q_v}{R_d/R_v}), 
    \\
\label{GovEqs_Supplement-c}
T
  & = \theta \left(\frac{p}{p_{\text{ref}}}\right)^{\frac{R_d}{c_{\text{pd}}}} \equiv \theta \pi, 
    \\
\label{GovEqs_Supplement-d}
\vect{v}
  & = \vect{u}+w\vect{k}, 
    \\
\label{GovEqs_Supplement-e}
S_{\text{ev}}
  & = C_{\text{ev}}\frac{p}{\rho}(q_{\text{vs}}-q_v)^+q_r, 
    \\
\label{GovEqs_Supplement-f}
S_{\text{cd}}
  & = C_{\text{cn}}(q_v-q_{\text{vs}})^+q_{\text{cn}}+C_{\text{cd}}(q_v-q_{\text{vs}})q_c, 
    \\
\label{GovEqs_Supplement-g}
S_{\text{ac}}
  & = C_{\text{ac}}(q_c-q_{\text{ac}})^+, 
    \\
\label{GovEqs_Supplement-h}
S_{\text{cr}}
  & = C_{\text{cr}}q_cq_r, 
    \\
\label{GovEqs_Supplement-i}
C 
  & = c_{\text{pd}}+c_{\text{pv}}q_v+c_l(q_c+q_r), 
    \\
\label{GovEqs_Supplement-j}
\Sigma
  & = (\frac{c_{\text{pv}}}{c_{\text{pd}}}R_d-R_v)q_v+\frac{c_l}{c_{\text{pd}}}R_d(q_c+q_r), 
    \\
\label{GovEqs_Supplement-k}
L(T) 
  & = L_{\text{ref}}-(c_l-c_{\text{pv}})(T-T_{\text{ref}})\equiv L_{\text{ref}}\phi(T).
\end{align}
\end{subequations}
This system extends \eqref{GovEqs_dry} by three transport equations for the moist constituents $q_v$, $q_c$ and $q_r$, denoting the mixing ratios of water vapor, cloud water and rain, respectively. As before, terms of the form $\mathcal{D}_\star$ denote generic closures for turbulent mixing and diffusion, while the terminal rainfall velocity $V_r$ appears as a new parameter. There is no universally valid scaling for this quantity, for reasons explained in the next section, and that is why we do not include it in Table \ref{tab:thermodynamics_moist} below. Notice further that the effect of moisture on total density is accounted for by Eq.~\eqref{GovEqs_Supplement-a}, where $\rho_d$ stands for the density of dry air. As far as the pressure is concerned, we assume ideal gas behavior for \emph{both} dry air and water vapor, implying in particular the relation
\begin{equation}
    e = R_v \rho_v T = R_v \rho_d q_v T
\end{equation}
for the vapor pressure $e = p - p_d$. Eq.~\eqref{GovEqs_Supplement-b} then results from adding the two ideal gas laws. Finally, the terms $S_{\text{ev}}$, $S_{\text{cd}}$, $S_{\text{ac}}$ and $S_{\text{cr}}$ denote closures for the microphysical processes evaporation, condensation, autoconversion of cloud droplets into raindrops, and collection of cloud water by falling rain, respectively. We defer a thorough discussion of this closure scheme and its integration into the asymptotic framework to the next section, only pointing out their dependence on the saturation mixing ratio of water vapor $q_{\text{vs}}$, which to leading order is determined by CC and therefore itself a function of temperature (see the discussion below).

Having thus set the stage, our point of departure is Clausius-Clapeyron (CC) in the form valid for an ideal gas (a very good approximation for atmospheric water vapor), which reads
\begin{equation} \label{CC_differential}
    \frac{d \ln e_s}{dT} = \frac{L(T)}{R_v T^2}.
\end{equation}
Here, $e_s = e_s(T)$ stands for the (nondimensional) saturation vapor pressure over a flat surface of liquid water at temperature $T$, while $R_v$ is the specific gas constant of water vapor, and $L(T)$ the enthalpy (latent heat) of vaporization, which also varies with temperature. For the latter, it is generally considered a valid approximation to assume a linear relationship, which reads
\begin{equation} \label{latent_heat_lin}
    L(T) = L_{\text{ref}} - (c_l - c_{\text{pv}})(T - T_{\text{ref}}).
\end{equation}
In this approximation, the specific heat capacities of water vapor at constant pressure and of liquid water are denoted by $c_{\text{pv}}$ and $c_l$, respectively, assumed constant. We should mention that the dependence of $c_l$ on temperature can be significant, but over the range of temperatures that we encounter in the lowest $\sim 15\;\text{km}$ of the atmosphere, the corresponding variations can be neglected.
\begin{remark}
    We do not include ice and mixed-phase clouds in our considerations. While these are clearly of great importance for a comprehensive understanding of moisture in the atmosphere, to the best of the authors' knowledge, the only systematic asymptotic analysis of atmospheric dynamics with explicit closures for the ice phase to date has been conducted by \citet{dolaptchiev2023}, who studied interactions between gravity wave dynamics and cirrus clouds, building on the application of asymptotic methods to the homogeneous nucleation process by \citet{baumgartner2019}. No fully general framework that treats both liquid and solid hydrometeors, coupled to the compressible flow equations, has been developed as of yet.
\end{remark}
Next, going back to the already established distinguished limit \eqref{dist_lim_dry}, we need to make one important adjustment: in Table \ref{tab:univ_char}, the isentropic exponent of dry air $\gamma$ was introduced, a dimensionless quantity. This parameter can be written in terms of $R_d$ and the specific heat capacity of dry air at constant pressure, $c_{\text{pd}}$. In the following, we replace $\gamma$ by $c_{\text{pd}}$ and reconsider the asymptotic rescaling of the ratio $R_d / c_{\text{pd}} = (\gamma - 1) / \gamma$, which under the standard assumption $\gamma = O(1)$ would also be a bounded quantity in the limit process $\epsilon \rightarrow 0$. The reason for this change will become apparent momentarily. - We thus consider the parameters listed in Table \ref{tab:thermodynamics_moist} below, with the scalings in \eqref{dist_lim_dry} still in place, and try to find a distinguished limit that enables us to represent moist processes in the earth's atmosphere in a physically and mathematically sound way.
\begin{table}[H]
    \centering
    \begin{tabular}{l|r}
       Dry air gas constant  & $R_d = 287$ J/kg/K \\
       Heat capacity of dry air at constant pressure  & $c_{\text{pd}} = 1005$ J/kg/K \\
       Water vapor gas constant & $R_v = 462$ J/kg/K \\
       Heat capacity of water vapor at constant pressure & $c_{\text{pv}} = 1850$ J/kg/K \\
       Heat capacity of liquid water & $c_l = 4218$ J/kg/K \\
       Enthalpy of vaporization at $T = T_{\text{ref}}$ & $L_{\text{ref}} \approx 2.5 \times 10^6$ J/kg
    \end{tabular}
    \caption{Fundamental thermodynamic parameters for a cloudy, ice-free atmosphere}
    \label{tab:thermodynamics_moist}
\end{table}
Now, before we proceed to discuss the derivation of an extended distinguished limit, let us state the three principles that shall guide us in this endeavor:
\begin{enumerate}
    \item \label{guideline1} Our aim is to contribute to the theoretical understanding of real-world phenomena. Therefore, we will only take scalings into consideration that lead to physically reasonable relations.
    \item \label{guideline2} As we now have twelve physical parameters and three nondimensional parameters have been assigned asymptotic rescalings already, the fundamental rules of dimensional analysis dictate that only five further scalings can be chosen freely.
    \item \label{guideline3} Recall that, in the sense of a heuristic correspondence, we assumed $\epsilon \sim 1/10$. As far as possible, we will assign scalings that reflect the numerical values of our thermodynamic parameters accordingly.
\end{enumerate}
As a first step, let us take a closer look at the thermodynamic evolution equation \eqref{GovEqs_temp}. Notwithstanding the great variation in cloud types and precipitation intensities across spatial and temporal scales, one statement holds true for \emph{all} atmospheric flow phenomena that are substantially influenced by moist processes: latent heating, i.e.~temperature changes connected to phase changes of atmospheric water, is present in the leading-order equation for spatiotemporal thermodynamic perturbations. This translates to
\begin{equation}
    C \left(D_t \tilde{\theta} + \tilde{w} \frac{d \bar{\theta}}{dz}\right) \sim \frac{L(\bar{T})}{\bar{T}} \left(D_t \tilde{q}_v + \tilde{w} \frac{d\bar{q}_{\text{vs}}}{dz}\right)
\end{equation}
in a generic formal asymptotic ansatz, where $\bar{f} = \bar{f}(z)$ is a given vertical background profile and $\tilde{f}$ denotes perturbations with unrestricted dependence on $(t,\vect{x},z)$ about a vertical background state. Depending on the cloud and / or precipitation type under consideration, one of the terms on either side may of course drop out of the leading-order balance. We thus see that the scaling of the nondimensional parameter
\begin{equation}
    \frac{L_{\text{ref}}}{c_{\text{pd}} T_{\text{ref}}} \approx 9.1
\end{equation}
plays a crucial role in determining the relative strength of moist thermodynamic perturbations. Here, the only sensible choice in accordance with guideline 3 is to assign the scaling
\begin{equation} \label{scaling_L}
    \frac{L_{\text{ref}}}{c_{\text{pd}} T_{\text{ref}}} = L \epsilon^{-1},
\end{equation}
which we shall treat as a given from now on. Next, going back to CC in differential form \eqref{CC_differential}, notice that the ansatz \eqref{latent_heat_lin} with constant heat capacities lets us integrate it exactly, leading to
\begin{equation} \label{es_exact}
    e_s(T)={e_s}_{\text{ref}}\left(\frac{T_{\text{ref}}}{T}\right)^{\frac{c_l-c_{\text{pv}}}{R_v}}\exp{\left\{\left(\frac{L_{\text{ref}}}{R_vT_{\text{ref}}}+\frac{c_l-c_{\text{pv}}}{R_v}\right)\frac{T - T_{\text{ref}}}{T}\right\}}.
\end{equation}
Since, by the ideal gas laws summarized in \eqref{GovEqs_Supplement-b}, $e_s$ is related to the saturation mixing ratio $q_{\text{vs}}$ by the formula
\begin{equation} \label{qvs_formula}
    q_{\text{vs}} = \frac{R_d}{R_v} \frac{e_s}{p - e_s},
\end{equation}
fixing the asymptotic magnitude of $e_s$ relative to the dry air pressure $p - e_s$ and determining a suitable scaling for $R_d / R_v$ will also fix the asymptotic magnitude of $q_{\text{vs}}$, which in turn provides an upper bound for the local water vapor mixing ratio $q_v$. This holds true because supersaturation attained in the real atmosphere is invariably quite small. Let us also bear in mind the following:
\begin{enumerate}
    \item[(i)] The saturation vapor pressure at $T = T_{\text{ref}}$ is very small compared to the total atmospheric pressure close to the surface.
    \item[(ii)] At leading order, temperature decreases with height in the troposphere.
    \item[(iii)] The exponential term makes the dominant contribution in \eqref{es_exact} when temperatures decrease, since $x^\alpha = e^{\alpha \ln x}\leq e^{\alpha (x  - 1)} \quad \forall x>0$, and hence
    \begin{equation}
        e_s(T) \leq \exp{\left\{\frac{L_{\text{ref}}}{R_v T_{\text{ref}}} \frac{T - T_{\text{ref}}}{T}\right\}}
    \end{equation}
    for all $T$.
\end{enumerate}
In determining an asymptotic representation of CC, describing the rapid decrease of available moisture at greater atmospheric altitudes, we thus need to look to the asymptotic expansion of the term
\begin{equation}
    \frac{L_{\text{ref}}}{R_v T_{\text{ref}}} \frac{T - T_{\text{ref}}}{T}.
\end{equation}
It is now straightforward to see why the standard dry air limit $R_d / c_{\text{pd}} = O(1)$ is troublesome when applied to CC: the resulting thermodynamic background state includes a linearly decreasing absolute temperature at leading order, implying
\begin{equation}
    \frac{T - T_{\text{ref}}}{T} = O(1).
\end{equation}
Moreover, we have
\begin{equation} \label{es_exponent}
    \frac{L_{\text{ref}}}{R_v T_{\text{ref}}} = \frac{R_v}{c_{\text{pd}}} \frac{L_{\text{ref}}}{c_{\text{pd}} T_{\text{ref}}} > \frac{L_{\text{ref}}}{c_{\text{pd}} T_{\text{ref}}},
\end{equation}
so this factor should by all rights be asymptotically \emph{large.} Adopting the standard limit now yields
\begin{equation}
    e_s(T) \sim \exp{\left(\frac{L_{\text{ref}}}{R_v T_{\text{ref}}} \frac{T - T_{\text{ref}}}{T}\right)} \sim \exp{\left(-\frac{1}{\epsilon}\right)},
\end{equation}
for temperatures $T  < T_{\text{ref}}$. The saturation vapor pressure consequently vanishes \emph{to all orders} above said temperature threshold, with translates to the statement that any nonzero amount of moisture at such temperatures (which are characteristic of middle to high tropospheric levels) \emph{immediately crosses the saturation threshold.} This is clearly unphysical because a significant amount of water vapor is present even in the driest regions of the real troposphere.
\begin{remark}
    The chosen reference temperature $T_{\text{ref}} = 273.15$ K is significantly lower than average surface temperatures in the tropics and most of the middle latitudes. Therefore, one should also be interested in the asymptotic behavior of CC for $T>T_{\text{ref}}$. Here, the situation is even more dire: the exponent \eqref{es_exponent} (which can be seen to determine the leading-order behavior also in this case) blows up, and the formal limit does not even exist in the first place.
\end{remark}
Going back to the drawing board, let us recall that the \emph{potential} temperature can generally be taken constant at leading order, since its tropospheric variation is of the order $\sim 30-40$ K. Due to the relationship
\begin{equation}
    T = \theta \left(\frac{p}{p_{\text{ref}}}\right)^{\frac{R_d}{c_{\text{pd}}}},
\end{equation}
absolute temperature will be constant at leading order if and only if $R_d / c_{\text{pd}} = o(1)$ in the adopted distinguished limit (pressure always exhibits $O(1)$ variations). To achieve a finite saturation vapor pressure in the limit $\epsilon \rightarrow 0$, one can thus propose the scaling
\begin{equation} \label{scaling_newtonian_lim}
    \frac{R_d}{c_{\text{pd}}} = \epsilon \Gamma,
\end{equation}
known as the \emph{Newtonian limit,} borrowed from combustion theory \citep{parkins2000}. This is the choice made by \citet{klein2006}, later also adopted by \citet{hittmeir2018}, as well as \citet{baeumer2025}. With the Newtonian limit in place, the exponent of the dominant term in \eqref{es_exact} reads
\begin{equation}
    \epsilon^{-2} \frac{R_d}{R_v} \frac{L}{\Gamma} \frac{T - T_{\text{ref}}}{T} \sim  O(\epsilon^{-1}) \frac{R_d}{R_v}.
\end{equation}
Thus, in order to achieve $O(1)$ variations of the saturation vapor pressure throughout the free troposphere in the formal limit, we must also assume
\begin{equation} \label{scaling_gas_const}
    \frac{R_d}{R_v} = \epsilon E,
\end{equation}
with $E = O(1)$ in the limit $\epsilon \rightarrow 0$. This scaling is still somewhat awkward when looking at the numerical value
\begin{equation}
    \frac{R_d}{R_v} \approx 0.62,
\end{equation}
but it is forced if we respect guidelines \ref{guideline1} and \ref{guideline2} \emph{and} try to do the least amount of violence to guideline \ref{guideline3}. Having established \eqref{scaling_L}, \eqref{scaling_newtonian_lim} and \eqref{scaling_gas_const}, only asymptotic scalings for the ratios of the various heat capacities remain to complete the description of the distinguished limit for a moist atmosphere. Here, we need to take a closer look at the coefficient \eqref{GovEqs_Supplement-j} in the thermodynamic equation. In scaled form, the term
\begin{equation} \label{scaling_Sigma}
    \frac{c_{\text{pv}}}{c_{\text{pd}}} \frac{R_d}{c_{\text{pd}}} - \frac{R_v}{c_{\text{pd}}} \approx 0.067
\end{equation}
only adjusts toward a positive value if we assume $c_{\text{pv}} / c_{\text{pd}} = O(\epsilon^{-1})$, which corresponds to a Newtonian limit for water vapor. On the other hand, the heat capacity of liquid water is significantly larger than that of water vapor,
and hence we choose
\begin{align}
    \frac{c_l}{c_{\text{pd}}} = \epsilon^{-1} k_l, & \quad \frac{c_{\text{pv}}}{c_{\text{pd}}} = \epsilon^{-1} k_v.
\end{align}
\paragraph{The empiricist's alternative:} \citet{hittmeir2018}, who first derived the above distinguished limit in the context of the dynamics of convective cloud towers, also considered an alternative approach that sacrifices formal consistency for more accurate representation of the actual numerical values. To be specific, they proposed a limit that deviates from the formally consistent one \emph{only} in the scalings of $R_d / R_v$ and $c_{\text{pv}} / c_{\text{pd}}$, both of which were considered $O(1)$ when occurring in isolation; in CC, nevertheless, the ratio $L_{\text{ref}} / (R_v T_{\text{ref}})$ was scaled as an $O(\epsilon^{-1})$ quantity, which leads to an obvious contradiction when writing it as a product (as in \eqref{es_exponent}). Similarly, the term \eqref{scaling_Sigma} was ad hoc assigned an asymptotic magnitude of $O(\epsilon)$, implying another formal inconsistency. Even so, the reduced models coming out of such an ansatz might have merit - in the words of the authors of the cited article:
\begin{quote}
    ``The second regime, in contrast, has been defined purely on the basis of the actual magnitudes of the dimensionless parameters. Numbers between $0.4$ and $3.0$ are considered of order unity, while smaller or larger values are associated with asymptotic rescalings in terms of $\epsilon$. This provides a scaling that better matches with the actual numbers than the first regime, but it is not strictly consistent with similarity theory. Although this is at odds with the usual procedures, it may actually open up an interesting route of investigation. The thermodynamics of moist air may just be asymptotically compatible with a family of equation systems that features the same functional forms in the constitutive equations as those of moist air, but whose set of determining parameters is less constrained. The results of Sect.~4.2, in which we compare asymptotic and error-controlled numerical approximations to the moist adiabatic distribution, corroborate this point of view.''
\end{quote}
Beyond \citet{hittmeir2018}, said limit was also adopted in the derivation of the large-scale model of \citet{baeumer2025}, mainly because it preserves the traditional form of the hydrostatic balance relation. The jury is still out on the question whether one of the two distinguished limits described here can generally be considered ``better'' than the other. Adopting the terminology of the two cited articles, we refer to the formally consistent regime as $\alpha = 0$, while the value-based alternative is indicated by $\alpha = 1$, see Table \ref{tab:regimes_moist} for an overview.
\begin{table}[H]
    \centering
    \begin{tabular}{l l l l}
    \hline
         & Value & Regime $\alpha = 1$ & Regime $\alpha = 0$ \\
         \hline
        \emph{Nondimensional parameters} & & & \\
        $\frac{R_d}{c_{\text{pd}}}$ & $0.29$ & $\epsilon \Gamma$ & $\epsilon \Gamma$ \\
        $\frac{c_{\text{pv}}}{c_{\text{pd}}}$ & $1.8$ & $k_v$ & $\epsilon^{-1}k_v$ \\
        $\frac{R_v}{c_{\text{pd}}}$ & $0.46$ & $1/A$ & $1/A$ \\
        $\frac{c_l}{c_{\text{pd}}}$ & $4.2$ & $\epsilon^{-1}k_l$ & $\epsilon^{-1}k_l$ \\
        $\frac{L_{\text{ref}}}{c_{\text{pd}}T_{\text{ref}}}$ & $9.1$ & $\epsilon^{-1}L$ & $\epsilon^{-1}L$ \\
        \emph{Derived nondimensional parameters} & & & \\
        $\frac{R_d}{R_v}$ & $0.62$ & $E$ & $\epsilon E$ \\
        $\frac{c_{\text{pv}}}{c_{\text{pd}}}\frac{R_d}{c_{\text{pd}}}-\frac{R_v}{c_{\text{pd}}}$ & $0.067$ & $\epsilon \kappa_v$ & $\kappa_v$ \\
        \hline
    \end{tabular}
    \caption{Two viable moist extensions of the distinguished limit for dry air}
    \label{tab:regimes_moist}
\end{table}
\paragraph{Scaling the moist constituents:} Mixing ratios are dimensionless quantities, and in the case of atmospheric water, they are invariably small. Even in the tropical troposphere, the biggest atmospheric moisture reservoir, the mixing ratio of water vapor at saturation does not exceed a couple percent. Therefore, the scaling
\begin{equation} \label{qvs_scaling}
    q_{\text{vs}} = O(\epsilon^2)
\end{equation}
suggests itself. Recalling \eqref{qvs_formula}, this implies that we scale the saturation vapor pressure as
\begin{equation}
   \frac{e_{s_{\text{ref}}}}{p_{\text{ref}}} = O(\epsilon^{1 + \alpha}),
\end{equation}
with $\alpha \in \{0,1\}$ referring to the two regimes summarized in Table \ref{tab:regimes_moist}. Again, the formally consistent regime $\alpha = 0$ can be seen to systematically \emph{overestimate} moist contributions relative to the numerical values. As already mentioned, supersaturation in the earth's atmosphere rarely exceeds even $1 \%$ \citep{houze2014}, which implies that \eqref{qvs_scaling} also sets the maximum magnitude of the water vapor mixing ratio $q_v$. Turning to the mass fractions of liquid water, it is not hard to see that no universally valid scaling can be expected: while the available supply of water vapor sets an obvious upper limit also for the magnitudes of $q_c$ and $q_r$, the phenomenologically ``correct'' scaling depends on the cloud type and therefore the scale under consideration. To be more specific, convective cloud towers as discussed, e.g., by \citet{hittmeir2018} produce much more intense rain than the stratiform clouds characteristic of midlatitude cyclones. Even typical terminal fall velocities vary significantly between the two phenomena, since convective updrafts produce much bigger raindrops. This is why the scalings of both the rain mixing ratio $q_r$ \emph{and} the terminal velocity $V_r$ differ between the cited study and the large-scale models of, e.g., \citet{smith2017} or \citet{baeumer2025}.
\section{Phase changes and their parameterization}
\subsection{Choosing an appropriate closure scheme}
The representation of phase changes of water on the macroscopic level of atmospheric motions poses one of the biggest challenges for both the theoretical and the numerical modeler: condensation kernels on which cloud droplets start to form have diameters on the order of \emph{microns.} As more and more liquid water diffuses onto a droplet, it may grow to precipitable size, gain a non-negligible fall speed, and then continue to accelerate downward until a balance between gravity and the drag force exerted on it by the surrounding air is reached. The droplet - now a raindrop - then attains its terminal velocity. This velocity depends on the drop's size, which in turn is a function of its interactions with other drops and droplets; crucially, it strongly increases with the intensity of local updrafts. Finally, as the raindrop gets closer to the ground, it may either fall unimpeded or (partially) evaporate in dry boundary layer air.

This sketch of the life cycle of one individual drop makes the complexity of the subject matter apparent: there is no clear-cut distinction between non-precipitating and precipitating water drops, fall speeds of the latter vary significantly, and they strongly depend on environmental conditions. Then, of course, there are the intricacies of the two-stage process of nucleation and condensation, which happen on very small spatial and temporal scales. In the hydrodynamic setting, we therefore need to assume quite drastic simplifications from the outset. In particular, in the context of asymptotic model hierarchies, it is prudent to keep the complexity of the chosen closure scheme to a manageable minimum. This is why the governing equations \eqref{GovEqs} include transport equations for only three moist constituents. Even though, as already remarked, the subdivision of liquid water into cloud water and rain is somewhat arbitrary, this simple bulk model is sufficient to capture the essentials of moist dynamics on all meteorological scales (without the ice phase). Phase changes in such a model are traditionally parameterized in a form originally devised by \citet{kessler1995}; for the reader's convenience, we restate the closure scheme employed here below. It goes back to \citet{klein2006}, and has further been utilized in the asymptotic modeling studies of \citet{hittmeir2018}, \citet{baeumer2023} and \citet{baeumer2025}:
\begin{subequations} \label{microphysics_closures}
    \begin{align}
S_{\text{ev}}
  & = C_{\text{ev}}\frac{p}{\rho}(q_{\text{vs}}-q_v)^+q_r, 
    \\
S_{\text{cd}}
  & = C_{\text{cn}}(q_v-q_{\text{vs}})^+q_{\text{cn}}+C_{\text{cd}}(q_v-q_{\text{vs}})q_c, 
  \label{microphysics_closures_cd}  \\
S_{\text{ac}}
  & = C_{\text{ac}}(q_c-q_{\text{ac}})^+, 
  \label{microphysics_closures_ac}  \\
S_{\text{cr}}
  & = C_{\text{cr}}q_cq_r.
    \end{align}
\end{subequations}
In the above relations, terms of the form $C_\star$ denote rate constants of the associated processes. The first term from the left in \eqref{microphysics_closures_cd} constitutes a closure for \emph{heterogeneous nucleation}, the initial formation of a droplet around a condensation nucleus; here, $q_{\text{cn}}$ represents the local density of cloud condensation kernels. In \eqref{microphysics_closures_ac}, $q_{\text{ac}}$ denotes an ``activation threshold'' for the conversion of cloud water into rain, as introduced by \citet{kessler1995}. We remark that the loss of smoothness at the phase interface makes the analytical treatment of the moisture equations \eqref{GovEqs_qv}-\eqref{GovEqs_qr} challenging, whether they are passively transported by a given velocity field or coupled to the full governing equations. See \citet{hittmeir2017,hittmeir2020,hittmeir2023} for global well-posedness results regarding the hydrostatic primitive equations and \citet{doppler2024} for a local result concerning the full compressible system \eqref{GovEqs}. All of the cited works utilize standard closures for turbulent diffusion and mixing for access to parabolic regularity.
\begin{remark}
    We emphasize that \eqref{microphysics_closures} was chosen pragmatically as the simplest qualitatively accurate bulk scheme, where all phase transitions are functions of the respective mixing ratios or their excess beyond a certain threshold only. More realistic bulk microphysics closures involve fractional powers of the liquid water mixing ratios, which are quite awkward to use in a formal asymptotic framework. For the reader interested in more accurate parameterizations, the overview in Chapter 3 of \citet{houze2014} is a good starting point. We further mention the double-moment scheme of \citet{seifert2001} as an example of a parameterization that balances moderate complexity with a systematic treatment of autoconversion, accretion and self-collection.
\end{remark}
\subsection{Asymptotic treatment of phase changes}
At first sight, it might seem straightforward to find proper scalings for the conversion terms \eqref{microphysics_closures} on the respective right-hand sides of the moisture equations \eqref{GovEqs_qv}-\eqref{GovEqs_qr}: having determined the appropriate asymptotic magnitudes of the various mixing ratios and the scaled value of the average terminal rainfall velocity $V_r$ for a given time - and length scale, one estimates the scaled conversion rates relative to the reference time and can thus determine their respective asymptotic strengths. The problem with this naive approach is that the rate constants on the respective right-hand sides of \eqref{microphysics_closures}, essentially representing ``best fits'' for long-time averages, will in general depend on the type of cloud and precipitation under consideration. Specifically, since the bulk modeling approach does not differentiate between drops and droplets of different sizes, we cannot expect all of the characteristic differences between convective and stratiform precipitation to be captured by the mixing ratios of cloud water and rain alone. Therefore, we certainly should not assume \emph{one} particular choice of the constants $C_\star$ to be valid across all meteorological scales, and instead critically re-evaluate the respective strengths of all microphysical processes for each scale-specific model. Awareness of this is of particular importance when looking ahead to multiscale models that will systematically explore interactions between moist processes from the convective to the synoptic scale. Now, going back to already established single-scale models, we can discern two principally different approaches to the modeling of phase transitions in asymptotic studies:
\begin{enumerate}
    \item \emph{Continuous re-parameterization:} This modeling strategy, pursued by, e.g., \citet{baeumer2023} and \citet{baeumer2025}, assumes that the averaged effects of the various microphysical processes on the given meteorological scale can be parametrized by continuous functions of the same type as the original closures in \eqref{microphysics_closures}. As laid out above, this does \emph{not} amount to a scaling assumption on the original rate constants, since those might themselves be scale-dependent.
    \item \emph{Fast microphysics:} This ansatz builds on the straightforward observation that microphysical processes are resolved on accordingly small scales, and can therefore be considered immediate on the much larger meteorological timescales. Most asymptotic studies dealing with a moist atmosphere that the authors are aware of treat the condensation term in this manner, with some extending it to all phase transitions, see, e.g., \citet{smith2017}.
\end{enumerate}
As shown below, the transition from the first approach to the second can be achieved - at least for the condensation term - by a straightforward rescaling of the appropriate rate constants. We devote the last section to a detailed comparison of the respective outcomes, choosing the precipitating quasi-geostrophic (PQG) model family as a representative example.

Let us consider the following situation: we are in the process of constructing a reduced single-scale model by a generic regular asymptotic expansion, and the scaling is already fully in place. However, we want to obtain a fast microphysics limit for the condensation term \emph{as a byproduct} of the formal derivation. In this context, the \emph{fast condensation limit} refers to the following alternative:
\begin{align}\label{cd_fast}
    \left\{\begin{array}{ll}
        q_v=q_{\text{vs}},\ \ \ q_c=q_t-q_{\text{vs}}-q_r & \text{in saturated air} \\
        q_v<q_{\text{vs}},\ \ \ q_c=0 & \text{in undersaturated air,}
    \end{array}\right.
\end{align}
with the total water content $q_t = q_v + q_c + q_r$. This alternative reduces the number of prognostic variables by one, since $q_c$ can be calculated explicitly from $q_t$ and $q_r$ at any given time (and $q_{\text{vs}}$, which in general is a function of temperature and therefore also part of the system's output):
\begin{equation}
    q_c = \max \left(q_t-q_{\text{vs}}-q_r, 0\right).
\end{equation}
Similarly, one gets
\begin{equation} \label{implicit_qv}
    q_v = \min \left(q_{\text{vs}}, q_t - q_r\right).
\end{equation}
On a formal level, this alternative can be recovered systematically simply by rescaling the respective nucleation and condensation rates appropriately: instead of assuming continuous reparameterization, we recognize both $C_{\text{cn}}$ and $C_{\text{cd}}$ as asymptotically fast with respect to the chosen timescale and make the ansatz
\begin{equation}
    C_{\text{cn}} t_{\text{ref}} \sim C_{\text{cd}} t_{\text{ref}} \sim \epsilon^{-n},
\end{equation}
for some sufficiently large $n$. In the scaled moisture equations, this yields the leading-order balance
\begin{align}
    S_{\text{cd}}^{(0)} = & C_{\text{cn}} (q_v^{(1)} - q_{\text{vs}}^{(1)})^+ q_{\text{cn}}^{(0)} \nonumber \\
    & + C_{\text{cd}} (q_v^{(1)} - q_{\text{vs}}^{(1)}) q_c^{(0)} = 0,
\end{align}
which, since $q_c^{(0)}$ and $q_{\text{cn}}^{(0)}$ cannot be negative, immediately leads to \eqref{cd_fast}.
\section{Precipitating quasi-geostrophic models}
\subsection{Fundamentals}
The classical quasi-geostrophic (QG) theory is of central importance in the mathematical analysis of both atmospheric and oceanic fluid dynamics \citep{pedlosky1987,vallis2017}. In particular, it captures the essential dynamics of the midlatitude atmosphere with its alternating cyclonic-anticyclonic structure, and its standard formulation highlights the importance of \emph{potential vorticity,} a fundamental quantity in all of meteorology \citep{ertel1942,hoskins1985}. In the following, we will use the notation
\begin{equation}
    D_t^\parallel := \partial_t + \vect{u} \cdot \nabla_\parallel
\end{equation}
for the material derivative with respect to the horizontal velocity field. To set the stage for the rest of this section, let us very briefly recap the equations and their main properties: starting with the governing equations \eqref{GovEqs_dry} for dry air, systematic scale analysis - which can be performed in the \citet{klein2010} framework - and asymptotic expansion lead to
\begin{equation} \label{QG_dry_geostrophy}
    f \vect{u} = \nabla_\parallel^\perp \tilde{\phi}
\end{equation}
as the leading-order horizontal momentum balance, where $f$ is the Coriolis parameter at the reference latitude and $\phi = \tilde{p} / \bar{\rho}$ denotes the pressure perturbation scaled by the background density. This relation is known as \emph{geostrophic balance.} The vertical momentum equation yields 
\begin{equation} \label{QG_dry_hydrostasy}
    \partial_z \tilde{\phi} = g\frac{\tilde{\theta}}{\theta_{\text{ref}}},
\end{equation}
which is known as \emph{hydrostatic balance.} The first-order contribution to horizontal momentum, combined with the mass continuity equation, can be written in the form
\begin{equation} \label{QG_dry_vorticity}
    D_t^\parallel \left[\zeta + \beta y\right] = \frac{f}{\bar{\rho}} \partial_z (\bar{\rho} \tilde{w}),
\end{equation}
where $\zeta = v_x - u_y$ denotes the vertical component of the vorticity, $\beta$ the variation of the Coriolis force with latitude, and $\tilde{w}$ the small geostrophic vertical velocity. Finally, the thermodynamic Eq.~\eqref{GovEqs_dry_temp} reads
\begin{equation} \label{QG_dry_temp}
    D_t^\parallel \tilde{\theta} + \tilde{w} \frac{d\bar{\theta}}{dz} = Q
\end{equation}
at leading order. Here, $d\bar{\theta} / dz > 0$ is assumed throughout, which translates to the atmosphere being \emph{stably stratified} in the large-scale mean.

Having collected \eqref{QG_dry_geostrophy}-\eqref{QG_dry_temp}, one can achieve a significant simplification of this system by eliminating the vertical velocity: adding the vorticity Eq.~\eqref{QG_dry_vorticity} and the vertical derivative of a suitable multiple of \eqref{QG_dry_temp} naturally leads to
\begin{equation}
    D_t^\parallel \text{PV} = \frac{f}{\bar{\rho}} \partial_z \left(\frac{\bar{\rho}Q}{d\bar{\theta} / dz}\right),
\end{equation}
the evolution equation for the \emph{QG potential vorticity} ($\text{PV}$), defined as
\begin{equation} \label{QG_dry_PV_transport}
    \text{PV} := \zeta + \beta y + \frac{f}{\bar{\rho}} \partial_z \left(\frac{\bar{\rho}\tilde{\theta}}{d\bar{\theta} / dz}\right).
\end{equation}
Substituting the geostrophic and hydrostatic balances \eqref{QG_dry_geostrophy} and \eqref{QG_dry_hydrostasy}, we get
\begin{equation} \label{QG_dry_PV_inversion}
    \frac{1}{f} \Delta_\parallel \tilde{\phi} + \frac{f}{\bar{\rho}} \partial_z \left(\frac{\bar{\rho}\partial_z\tilde{\phi}}{N^2}\right) = \text{PV} - \beta y,
\end{equation}
where $N$ denotes the background buoyancy frequency,
\begin{equation}
    N := \sqrt{g \frac{d\bar{\theta} / dz}{\theta_{\text{ref}}}}.
\end{equation}
Due to the assumption of stable stratification stated above, $N$ is real-valued and \eqref{QG_dry_PV_inversion} constitutes a linear elliptic equation for $\tilde{\phi}$ when PV is given. The process of retrieving the pressure from \eqref{QG_dry_PV_inversion} is known as \emph{potential vorticity inversion.} Thus, the evolution of the QG system is governed exclusively be the PV transport Eq.~\eqref{QG_dry_PV_transport}, an \emph{active scalar equation.}

There is a rich mathematical literature on the QG equations. Global well-posedness of the full system is addressed, e.g., by \citet{bourgeois1994}, while \citet{novack2018,novack2020} cover its extension to a boundary layer theory. We would be remiss not to mention that the 2D \emph{surface QG} (SQG) equation, derived from a simplified version of the original system with $\text{PV} \equiv 0$, has captured the attention of mathematicians ever since \citet{constantin1994} alerted the mathematical fluid dynamics community to its intriguing parallels to the 3D Euler equations.
\subsection{Moist extension and the structural effect of fast microphysics}
The passage from continuous phase transitions to switches as in the ``fast condensation limit'' \eqref{cd_fast} can fundamentally alter the analytical structure of the resulting reduced model. The \emph{precipitating QG} (PQG) model family, going back to \citet{smith2017}, here serves as an ideal example, since variants of the PQG equations both with fast phase changes and with full Kessler-style closures have already been derived and investigated. We begin with the PQG model recently derived by \citet{baeumer2025}, which preserves the original form of the bulk microphysics closures \eqref{microphysics_closures}. In their ``raw'' form, the equations of this model in dimensional form read
\begin{subequations}\label{PQG_preliminary}
    \begin{align}
        f\vect{k}\times\vect{u}&=-\nabla_\parallel{\tilde{\phi}} \\
        \partial_z\tilde{\phi} &= g \frac{\tilde{\theta}}{\theta_{\text{ref}}} \\
        D_t^\parallel\left[\zeta+\beta y\right]&=\frac{f}{\bar{\rho}}\partial_z(\bar{\rho}\tilde{w}) \\
        \label{PQG_preliminary_theta_e}
        D_t^\parallel\tilde{\theta}_e+\tilde{w}\frac{d\bar{\theta}_e}{dz}&=0 \\
        \label{PQG_preliminary_qv}
        D_t^\parallel \tilde{q}_v+\tilde{w}\frac{d\bar{q}_{\text{vs}}}{dz}&=S_{\text{ev}}-S_{\text{cd}} \\
        \label{PQG_preliminary_qc}
        D_t^\parallel q_c &= S_{\text{cd}}-S_{\text{ac}}-S_{\text{cr}} \\
        \label{PQG_preliminary_qr}
        -\frac{1}{\bar{\rho}}\partial_z(\bar{\rho} V_r q_r)&=S_{\text{ac}}+S_{\text{cr}}-S_{\text{ev}},
    \end{align}
\end{subequations}
where $\theta_e = \theta + \frac{L_{\text{ref}}}{c_{\text{pd}}} q_v$ is the (linearized) equivalent potential temperature, $\tilde{q}_{\text{vs}}$ is a linear function of potential temperature $\tilde{\theta}$, by asymptotic expansion of CC, and the leading-order phase changes read
\begin{subequations}
    \begin{align}
        S_{\text{ev}} &= \bar{C}_{\text{ev}} \left(\tilde{q}_{\text{vs}}-\tilde{q}_v\right)^+ q_r \\
        S_{\text{cd}} &= \bar{C}_{\text{cn}} \left(\tilde{q}_v-\tilde{q}_{\text{vs}}\right)^+ q_{\text{cn}} + \bar{C}_{\text{cd}} \left(\tilde{q}_v-\tilde{q}_{\text{vs}}\right) q_c \\
        S_{\text{ac}} &= \bar{C}_{\text{ac}} \left(q_c-q_{\text{ac}}\right)^+ \\
        S_{\text{cr}} &= \bar{C}_{\text{cr}} q_c q_r.
    \end{align}
\end{subequations}
The derivation of a PV equation here proceeds in essentially the same fashion as for the classical dry air model. Due to the presence of a moist background, however, we further need to introduce a new moisture variable to fully eliminate the vertical velocity. We refer to \citet{baeumer2025} for a detailed derivation, only stating the result:
\begin{subequations} \label{PQG_DL_dim}
    \begin{align}
        D_t^\parallel \text{PV}_e =& -\frac{f}{d\bar{\theta}_e/dz} \partial_z \vect{u} \cdot \nabla_\parallel \left(\frac{L_{\text{ref}}}{c_{\text{pd}}}\tilde{q}_v\right) \\
        D_t^\parallel \tilde{M} =& B(z) \left[S_{\text{ev}} - S_{\text{cd}}\right] \\
        D_t^\parallel q_c =& S_{\text{cd}} - S_{\text{ac}} - S_{\text{cr}} \\
        -\frac{1}{\bar{\rho}} \partial_z \left(\bar{\rho} V_r q_r\right) =& S_{\text{ac}} + S_{\text{cr}} - S_{\text{ev}} \\
        \frac{1}{f}\Delta_\parallel \tilde{\phi} + \frac{f}{\bar{\rho}} \partial_z \left(\frac{\bar{\rho}}{d\bar{\theta}_e/dz}\frac{B(z)}{L_{\text{ref}}/c_{\text{pd}}+B(z)} \partial_z \tilde{\phi}\right) =& \text{PV}_e - \beta y \label{PQG_DL_dim_inversion} \nonumber \\
        &- \frac{f}{\bar{\rho}} \partial_z \left(\frac{\bar{\rho}}{d\bar{\theta}_e/dz}\frac{L_{\text{ref}}/c_{\text{pd}}}{L_{\text{ref}}/c_{\text{pd}}+B(z)} \tilde{M}\right) \\
        f \vect{k} \times \vect{u} =& -\nabla_\parallel \tilde{\phi} \\
        \partial_z \tilde{\phi} =& g\frac{\tilde{\theta}}{\theta_{\text{ref}}},
    \end{align}
\end{subequations}
where $\text{PV}_e$ is the QG potential vorticity based on linearized equivalent potential temperature $\theta_e = \theta + \frac{L_{\text{ref}}}{c_{\text{pd}}} q_v$, the moisture variable $\tilde{M}$ is a linear combination of $\tilde{\theta}_e$ and $\tilde{q}_v$ and the following additional relations hold:
\begin{subequations}
    \begin{align}
        B(z) &= -\frac{d\bar{\theta}_e/dz}{d\bar{q}_{\text{vs}}/dz} \\
        \tilde{q}_v &= \frac{1}{L_{\text{ref}}/c_{\text{pd}}+B(z)} \left(\tilde{M} - \tilde{\theta}\right).
    \end{align}
\end{subequations}
Notice that the $\text{PV}_e$ inversion equation here is still linear when all prognostic variables are given.
\subsection{$\text{PQG}_{\text{DL}}$ in the fast condensation limit}
Let us now go back to the reduced moisture and thermodynamic equations in their preliminary form to examine the effect of fast condensation: we have
\begin{subequations}
    \begin{align}
        D_t^\parallel \tilde{q}_v + \tilde{w} \frac{d \bar{q}_{\text{vs}}}{dz} &= S_{\text{ev}} - S_{\text{cd}}^{(n)} \\
        D_t^\parallel q_c &= S_{\text{cd}}^{(n)} - S_{\text{ac}} - S_{\text{cr}} \\
        -\frac{1}{\bar{\rho}} \partial_z (V_r \bar{\rho} q_r) &= S_{\text{ac}} + S_{\text{cr}} - S_{\text{ev}} \\
        D_t^\parallel \tilde{\theta}_e + \tilde{w} \frac{d\bar{\theta}_e}{dz} &= 0,
    \end{align}
\end{subequations}
which necessitates elimination of the higher-order term $S_{\text{cd}}^{(n)}$. This leads to the introduction of the auxiliary variable $\tilde{q}_{\text{vc}} = \tilde{q}_v + q_c$ and the reduced system
\begin{subequations}
    \begin{align}
        D_t^\parallel \tilde{q}_{\text{vc}} + \tilde{w} \frac{d\bar{q}_{\text{vs}}}{dz} &= \frac{1}{\bar{\rho}} \partial_z (V_r \bar{\rho} q_r) \\
        -\frac{1}{\bar{\rho}} \partial_z (V_r \bar{\rho} q_r) &= S_{\text{ac}} + S_{\text{cr}} - S_{\text{ev}} \\
        D_t^\parallel \tilde{\theta}_e + \tilde{w} \frac{d\bar{\theta}_e}{dz} &= 0.
    \end{align}
\end{subequations}
Eliminating the vertical velocity and deriving the $\text{PV}_e$ equation as before then yields
\begin{subequations}
    \begin{align}
        D_t^\parallel \tilde{M} &= \frac{B(z)}{\bar{\rho}} \partial_z (V_r \bar{\rho} q_r)  \\
        -\frac{1}{\bar{\rho}} \partial_z (V_r \bar{\rho} q_r) &= S_{\text{ac}} + S_{\text{cr}} - S_{\text{ev}} \\
        D_t^\parallel \text{PV}_e &= -\frac{f}{d\bar{\theta}_e / dz} \partial_z \vect{u} \cdot\nabla_\parallel \left(\frac{L_{\text{ref}}}{c_{\text{pd}}} \tilde{q}_v\right),
    \end{align}
\end{subequations}
where we can write
\begin{equation} \label{qv_implicit}
    \tilde{q}_v = \min \left(\tilde{q}_{\text{vs}},\frac{1}{L_{\text{ref}} / c_{\text{pd}} + B(z)}(\tilde{M} - \tilde{\theta)}\right)
\end{equation}
($\tilde{M}$ is now defined in terms of $\tilde{\theta}_e$ and $q_{\text{vc}}$). Determining the local saturation status therefore requires knowledge of $\tilde{\theta}$, which is clearly part of the \emph{output} of the $\text{PV}_e$ inversion equation. In said equation,
\begin{equation}
    \frac{1}{f} \Delta_\parallel \tilde{\phi} + \frac{f}{\bar{\rho}} \partial_z \left(\frac{\bar{\rho}\tilde{\theta}_e}{d\bar{\theta}_e / dz}\right) = \text{PV}_e - \beta y,
\end{equation}
we similarly can only write
\begin{equation} \label{theta_e_implicit}
    \tilde{\theta}_e = \min \left(\tilde{\theta} + \frac{L_{\text{ref}}}{c_{\text{pd}}} \tilde{q}_{\text{vs}}, \frac{B(z)}{L_{\text{ref}} / c_{\text{pd}} + B(z)} \tilde{\theta} + \frac{L_{\text{ref}} / c_{\text{pd}}}{L_{\text{ref}} / c_{\text{pd}} + B(z)} \tilde{M}\right).
\end{equation}
Therefore, the functional form of the inversion equation is only known on either side of the phase interface, which in turn depends on the solution of the equation. We thus get an a priori only \emph{piecewise} elliptic equation that involves a free boundary problem; this makes the analytical and numerical study of PQG models with fast microphysics very challenging. A first step toward the rigorous validation of such models was taken by \citet{remond-tiedrez2024}, who proved the well-posedness of the nonlinear PV inversion equation of \citet{smith2017} in a weak formulation. To summarize: we have obtained the system
\begin{subequations} \label{PQG_DL_fast}
    \begin{align}
        D_t^\parallel \text{PV}_e =& -\frac{f}{d\bar{\theta}_e/dz} \partial_z \vect{u} \cdot \nabla_\parallel \left(\frac{L_{\text{ref}}}{c_{\text{pd}}}\tilde{q}_v\right) \\
        D_t^\parallel \tilde{M} &= \frac{B(z)}{\bar{\rho}} \partial_z (V_r \bar{\rho} q_r) \\
        -\frac{1}{\bar{\rho}} \partial_z \left(\bar{\rho} V_r q_r\right) =& S_{\text{ac}} + S_{\text{cr}} - S_{\text{ev}} \\
         \frac{1}{f} \Delta_\parallel \tilde{\phi} + \frac{f}{\bar{\rho}} \partial_z \left(\frac{\bar{\rho}\tilde{\theta}_e}{d\bar{\theta}_e / dz}\right) &= \text{PV}_e - \beta y \label{PQG_DL_fast_inversion} \\
        f \vect{k} \times \vect{u} =& -\nabla_\parallel \tilde{\phi} \\
        \partial_z \tilde{\phi} =& g\frac{\tilde{\theta}}{\theta_{\text{ref}}},
    \end{align}
\end{subequations}
with the auxiliary relations \eqref{qv_implicit} and \eqref{theta_e_implicit}, as the fast condensation equivalent of \eqref{PQG_DL_dim}.
\section{Summary and outlook}
We discussed recent steps towards a general framework for the asymptotic modeling of moist atmospheric flows, omitting the ice phase for the time being. In particular, we showed in detail that the incorporation of the fundamental Clausius-Clapeyron relation makes it difficult to obtain a distinguished limit that is both formally consistent and faithful to the numerical magnitudes of the thermodynamic parameters of moist air. We argued, as \citet{hittmeir2018} did, that two different regimes could be considered viable, summarized in Table \ref{tab:regimes_moist}. Furthermore, we laid out two contrasting approaches to the modeling of phase changes and highlighted the profound structural changes arising from a transition to fast microphysics in the family of precipitating quasi-geostrophic models.

As far as directions for future research are concerned, we can only offer a personal selection, since the field has a whole is still in its infancy: on the modeling side, the systematic study of interactions between small-scale moist convection and large-scale cloud regions stands out as a problem of great theoretical and practical relevance. The connection of reduced models for moist atmospheric dynamics to the planetary boundary layer constitutes another important challenge. Here, \citet{baeumer2025} achieved significant progress by merging the recent triple-deck boundary layer theory of \citet{klein2022} with the PQG model family (the equations for the bulk flow are stated in \eqref{PQG_DL_dim}). Of course, much still needs to be learned about the properties of previously derived models. - Again highlighting developments in the PQG context, numerical convergence studies as in \citep{zhang2022} are a promising avenue of investigation, as is the derivation of families of physically meaningful explicit solutions, achieved by \citet{wetzel2019} for discontinuous fronts. Regarding the rigorous analysis of models based on the Navier-Stokes or Euler equations with moist process closures as in \eqref{GovEqs}, most research to date has been concerned with the well-posedness of the full system with viscosity and closures for turbulent diffusion and mixing. No results for the inviscid system seem to be available. Looking to reduced models, very few investigations have been conducted: besides the aforementioned work of \citet{remond-tiedrez2024}, the only rigorous results that the authors are aware of have been obtained by \citet{majda2010} and \citet{li2016} in the context of the early \citet{frierson2004} model for a moist tropical atmosphere. Remarkably, \citet{li2016} went beyond well-posedness and proved convergence to the so-called relaxation limit, which in our terminology corresponds to the transition to a fast microphysics scheme. It would be interesting to see whether results in the same spirit can be obtained for more comprehensive bulk models as in \eqref{microphysics_closures}. We should also note that one of the authors of the present article \citep{baeumer2025b} has obtained first results on the dry version of the triple-deck boundary layer theory developed by \citet{klein2022}. Last, but certainly not least, the question under which conditions and in which sense solutions of the unapproximated Eqs.~\eqref{GovEqs} converge to solutions of the reduced moist flow model under consideration is completely open - it remains to be seen whether a general framework in analogy to the classical theory of singular limits \citep{klainerman1981,schochet1994} can be developed.
\section*{Acknowledgements}
This research was funded in whole or in part by the Austrian Science Fund (FWF) 10.55776/F65. For open-access purposes, the authors have applied a CC BY public copyright license to any author-accepted manuscript version arising from this submission. R.K.~acknowledges support by Deutsche Forschungsgemeinschaft through Grant CRC 1114 ``Scaling Cascades in Complex Systems'', Project Number 235221301, Project C06 ``Multi-scale structure of atmospheric vortices'' and Grant FOR 5528 ``Mathematical Study of Geophysical Flow Models:
Analysis and Computation'', Project No.~500072446, Project~2 ``Scale Analysis and Asymptotic Reduced Models for the Atmosphere''. The authors thank the Wolfgang Pauli Institute Vienna for all kinds of support, e.g., the Pauli fellowship for R.K.
\bibliographystyle{SageH}
\bibliography{references}
\end{document}